\documentstyle[epsfig]{elsart}
\begin{document}

\begin{flushright}{NT@UW-03-08\\ FIU-PHY-050803\\SLAC-PUB-9854 }
\end{flushright}
\medskip
\begin{frontmatter}
\title{ Hard Photodisintegration of a Proton Pair in $^3$He}

\author{S.J.~Brodsky$^{1,2}$, L.~Frankfurt$^3$, R.~Gilman$^{2,4}$, 
J.R.~Hiller$^5$}

\author{\ \ \ \ \ G.A.~Miller$^6$, 
E.~Piasetzky$^3$, M.~Sargsian$^7$, M.~Strikman$^{8}$}

\address{
$^1$SLAC, Stanford University, Stanford, CA 94309 \\
$^2$Thomas Jefferson National Accelerator Facility, Newport News, VA 23606 USA \\
$^3$School of Physics and Astronomy, Sackler Faculty of Exact Science,
Tel Aviv University, Tel Aviv 69978, Israel\\
$^4$Rutgers University, Piscataway, NJ 08854 USA \\
$^5$University of Minnesota-Duluth, Duluth, MN 55812 \\
$^6$University of Washington, Seattle, WA 98195 \\
$^7$Florida International University, Miami, FL 33199 USA \\
$^8$Pennsylvania State University, University Park, PA 16802 USA}

\date{\today}

\begin{abstract}
Hard photodisintegration of the deuteron has been extensively studied
in order to understand the dynamics of the transition from 
hadronic to quark-gluon descriptions of the strong interaction. 
In this work, we discuss the extension of this program to
hard photodisintegration of a $pp$ pair in the $^3$He nucleus.
Experimental confirmation of new features 
predicted here for the suggested reaction would 
advance our understanding of hard nuclear reactions. 
A main prediction, in contrast with low-energy observations,
is that the $pp$ breakup cross section is not much smaller
than the one for $pn$ break up.
In some models, the energy-dependent oscillations observed for $pp$ scattering 
are predicted to appear in the $\gamma \; ^3{\rm He} \to p p + n$ 
reaction. Such an observation would open up a completely new field in studies 
of color coherence phenomena in hard nuclear reactions.
We also demonstrate that, in addition to the energy dependence, the
measurement of the light-cone momentum distribution of the recoil neutron  
provides an independent test of the underlying dynamics of hard 
disintegration.
\end{abstract}

\begin{keyword}
QCD, Hard Reactions, Photodisintegration
\PACS 24.85.+p,  25.10.+s, 25.20.-x, 25.70.Bc
\end{keyword}

\end{frontmatter}

\section{Introduction}
We define the hard photodisintegration of a nucleon pair as a process
in which a high energy photon is absorbed by a nucleon pair 
leading to  pair disintegration into  two nucleons with 
transverse momenta  greater than about 1 GeV/c.
In this process the Mandelstam parameters
$s$, the square of the total energy in the c.m.\ frame, and $t \approx u$,
the four-momentum transfers from the photon to the nucleons,
are large.  With $s$ above the resonance region, and $-t,-u$ $\ge$ 1 GeV$^2$, 
the kinematics are in the transition region, in which
the short distance scales probed might make it appropriate
to formulate the theory in terms of quark and gluon rather than hadronic  
degrees of freedom.

High-energy photodisintegration of a nucleon pair provides
an efficient way to reach the hard regime. To obtain the same $s$ in $NN$ 
scattering, one needs an incident nucleon lab momentum about a factor of two 
larger than that of the photon. Photodisintegration of a $pn$ pair, the 
deuteron, has now been extensively measured at high energies
\cite{NE8,NE17,crawford96,E89012,schulte01,schulte02,rossi02,wijesooriya01}. 
In this work, we investigate a related process,
the hard photodisintegration of a $pp$ pair, in the $^3$He nucleus. 

\begin{figure}[th]
\centerline{\epsfxsize=7.75cm \epsfbox{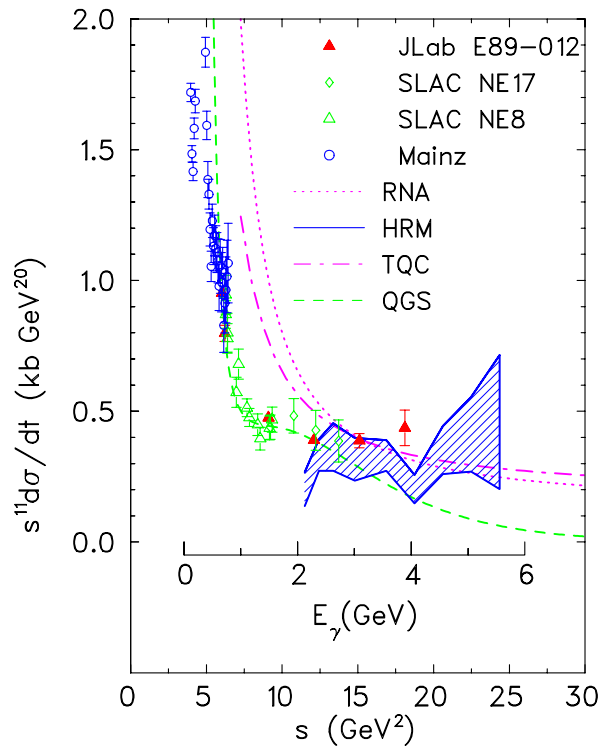} }
\caption{\it The energy dependence of 
$s^{11} {d\sigma\over dt}$ for
90$\,^{\circ}$ c.m.\ photodisintegration of the deuteron.
The HRM result is shown as a shaded band.
The QGS calculation is the long dashed line.  
The RNA result is the dotted line, normalized to 
the data point at 3 GeV.
The dot dash line shows the TQC formula, normalized to the 3 GeV data point.
The experimental data is labeled by the laboratory and the 
experiment number.
\label{gdpn}}
\end{figure}

Deuteron photodisintegration cross sections are available
for photon energies up to 5~GeV (but only 4~GeV at 
$\theta_{\rm c.m.}=90^{\circ}$)
\cite{NE8,NE17,crawford96,E89012,schulte01} 
including, for energies up to 2.5 GeV,
``complete'' angular distributions \cite{schulte02,rossi02} and 
recoil polarizations \cite{wijesooriya01}.
Figure~\ref{gdpn} shows the measured energy dependence of 
$s^{11} {d\sigma\over dt}$ for 90$^{\circ}$ c.m.
The quark counting rule prediction \cite{kn:it2,kn:it3,Polchinski:2001tt},
that this quantity becomes independent of energy,
is observed clearly.
High-energy deuteron photodisintegration cross sections 
at other angles are also 
in good agreement with scaling once $p_T$ $\ge$ 1.3 GeV/c. 

The good agreement of the data with the quark counting rule prediction
contrasts with observations \cite{Isgur_Smith,Radyushkin} that 
pQCD underestimates cross sections for
intermediate energy photo-reactions -- examples include the deuteron 
elastic form factor \cite{Farrar}, meson photoproduction \cite{Farrar2} 
and real Compton scattering \cite{BDixon}. 
Thus, it seems that although the observation of the scaling in a 
given reaction indicates the onset of the quark-gluon degrees of freedom, 
the appropriate underlying physics has a mixture of perturbative and 
nonperturbative QCD aspects.
A variety of theoretical models exist for deuteron photodisintegration
which explicitly account for quark-gluon degrees of freedom in the reaction 
with an  attempt to incorporate the nonperturbative QCD 
effects.\footnote{Note that to date there are no successful meson-baryon 
calculations
for the high energy data. For a recent review, see \cite{gilmangross}.}
Hidden color degrees of freedom of the 
nucleus might play an important role in determining the normalization of
hard-scattering nuclear amplitudes \cite{Farrar,Brodsky:1983vf}.

The reduced nuclear amplitude (RNA) formalism \cite{brodskyhiller} 
attempts to incorporate some of the soft physics not described by pQCD by using experimentally 
determined nucleon form factors to describe the gluon exchanges within the nucleons.
It neglects diagrams in which gluon exchanges between the nucleons lead
to non-color singlet intermediate ``nucleon'' states, diagrams which
might be important in pQCD calculations.
Ideally, the RNA calculation should be normalized
to the scaling behavior at asymptotic energies,
where both yield the same result.  In practice,
the normalization must be to data, but at energies
sufficiently large.  An estimate of the necessary
photon lab energy is obtained by requiring the
momentum transfer to each nucleon to be above
1 GeV, which yields \cite{CHH97}
\begin{equation}
\frac{1}{2} M_d E_\gamma 
  \left[ 1-\sqrt{\frac{2E_\gamma}{M_d+2E_\gamma}}
                  |\cos\theta_{\rm c.m.}|\right]
    \geq 1\,\mbox{GeV}^2\,.
\end{equation}

The two-quark coupling (TQC) model \cite{radyushkinmodel} is 
based on the idea that the photon interacts with a pair of quarks
being interchanged between the two nucleons.
An analysis of this hard interaction then shows that the
reaction has leading kinematic dependences proportional to
nucleon form factors, taken to be
the dipole form factor, to the fourth power
times a phase space factor times a propagator, $(s-\Lambda^2)^{-1}$,
where $\Lambda$ $\approx$ 1 GeV.
There is no absolute normalization predicted by the model;
instead it is normalized to the data at one point.
The formula manages to largely reproduce the energy and angle dependences 
of hard deuteron photodisintegration, for $E_{\gamma}$ $>$ 2 GeV,
once this one normalization parameter is fixed.
With the propagator $(s-\Lambda^2)^{-1}$, instead of the factor $p_T^{-2}$ 
in the similar RNA formula, the energy and angle dependences are softened,
improving the agreement with the data.

The quark-gluon string model (QGS) \cite{qgsmodel} views the reaction
as proceeding through three-quark exchange, with an
arbitrary number of gluon exchanges.
The cross section is evaluated using Regge theory techniques, and is
sensitive to the Regge trajectory used. 
While Regge theory has been shown to be an
efficient description of high-energy, small-$t$ reactions, it has
not typically been applied to the large momentum transfers being
discussed in this article.
The best fit of the data is 
obtained in a calculation that uses a nonlinear trajectory, as opposed 
to the more familiar linear trajectory.

The QCD hard rescattering model (HRM) \cite{nmisk} assumes 
that the photon is absorbed by a quark in one nucleon, followed by 
a high momentum transfer interaction with a quark of the other nucleon
leading to the production of  two nucleons with high relative momentum. 
Summing the relevant quark rescattering diagrams demonstrates that the 
nuclear scattering amplitude can be expressed as a convolution of the large 
angle $pn$ scattering amplitude, the hard photon-quark interaction vertex and 
the low-momentum  nuclear wave function. Since the $pn$ hard scattering 
amplitude can be taken from large angle $pn$ scattering data, 
the HRM model allows calculation of the absolute cross section of the 
$\gamma d \to pn$ reactions using no adjustable parameters.

Figure~\ref{gdpn} demonstrates the comparison of 
the calculations based on the models discussed above with the 
available data for deuteron disintegration at $\theta_{\rm c.m.}$
$=$ 90$^{\circ}$. RNA, TQC and QGS calculations require 
normalization to the data. The HRM does not require such a normalization 
factor, however the poor  accuracy  of hard-scattering $pn$ 
data restricts the overall accuracy of the calculation  to the level of 20\% --
this is shown as an error band in the figure. Each of the models describes 
some part of the data, but no model describes all of the data. 
Therefore further studies  to advance our understanding of  
hard photodisintegration reaction dynamics are needed.\footnote{ We also
note a recent study of deuteron photodisintegration in a
constituent quark model \cite{jdl02}.}

\section{Breaking a $pp$ Pair}
In the present work we suggest a new venue for studying  the 
dynamics of hard nuclear reactions. We propose to 
extend the studies of hard photodisintegration reactions
from the $pn$ system of the deuteron to the $pp$ system.
Namely, we propose the investigation of the reaction 
$\gamma \; ^3{\rm He} \to pp + n$
in which we define the measurement conditions so that the neutron in 
$^3$He can be considered, at least approximately,
as a static spectator, while two protons are produced at
$90^{\circ}$ in the c.m.\ frame of the $\gamma p p$ 
system.\footnote{ This can be done experimentally by selecting events in 
which the reconstructed missing neutron momentum is less than 100 MeV/c.}

The {\em advantage} of this program is that although many of the 
considered models do not predict the absolute cross section, still they 
can predict the relative cross section of the hard 
$\gamma (pp) \to pp$ reaction as compared to the 
$\gamma (pn) \to pn$ reaction. 
The $pn$ data from the deuteron already exist, and will be used in
this article to provide an overall normalization so that absolute
$\gamma ^3$He cross sections, rather than just the $s$ dependence of the
$\gamma pp$ cross section, can be predicted. 
The  nucleus   $^3$He  has been used successfully to observe
the absorption reaction $\pi^-pp\to np$\cite{ashery} at much lower
energies than appear here. Thus the use of $^3$He as a source of a $pp$ 
target has a successful 
history.\footnote{Measurements of the $pn$ break up in 
$^3$He are also possible, and would remove some uncertainty in the
nuclear physics aspects of the calculation.
For example, sensitivity to the high momentum component of the nuclear 
wave function would be reduced.}

\medskip

\noindent {\bf RNA model:}. 
In the RNA approach \cite{brodskyhiller}, the differential cross section is
proportional to the squares of form factors, one for each
nucleon, evaluated at the momentum transfer for that nucleon
in the weak-binding limit.  The remainder, the ``reduced''
cross section, is assumed to be independent of the substructure
of the nucleons.  This gives
\begin{equation}
\frac{d\sigma}{dt}\simeq F_{N_1}^2(-t_1) F_{N_2}^2(-t_2)
           \left.\frac{d\sigma}{dt}\right|_{\rm reduced}\,,
\end{equation}
for the process $\gamma (N_1 N_2)\to N_1 N_2$,
with $t_i$ the square of the four-momentum transfer to
nucleon $N_i$.  The ratio of cross sections for $\gamma(pp)\to pp$
and $\gamma(pn)\to pn$ is then given by the ratio of nucleon form
factors squared, $F_p^2(-t_N)/F_n^2(-t_N)$ ($t_N \approx {t\over 2}$), 
times the ratio of the reduced cross sections.  
The ratio of form factors can be obtained
from data for $G_M$ and $G_E$~\cite{BostedLung}; we use the leading
twist form factor $F_1$ for each nucleon, for which the ratio
$F_{1p}/F_{1n}$ is approximately \mbox{-2}.  The ratio of reduced
cross sections is taken to be 4, the square of the charge ratio.
These estimations yield $\gamma(pp)\to pp$ cross section  
approximately $16$ times larger than  the RNA prediction for 
$\gamma d\to pn$ cross section.
The absolute  normalization for the $\sigma_{RNA}(\gamma(pp)\to pp)$ 
can be obtained from comparison of $\sigma_{RNA}(\gamma d\to pn)$ with 
available data.

To estimate the cross section of $\gamma \; ^3{\rm He} \to pp + n$,
we shall multiply the above estimates of the
cross section of the disintegration of the $pp$ pair, 
$\sigma (\gamma (pp) \to pp)$, 
by a factor that combines 
the relative probability of a $pp$ pair in the $^3$He wave function
with a correction from the integration over the slow neutron's momentum. 
Note that no new normalization with the experimental data is needed, 
since we use the normalization factors obtained from the comparison of the 
$\gamma d \to p n$ cross sections with the data.

To estimate this factor we observe that in RNA the amplitude results from 
the $pp$ wave function at small separations. Therefore, as a simple estimate 
we use the parameter $a_2(A)$ which characterizes 
the probability of two-nucleon correlations in the nuclear wave function --
$a_2(A=3)\approx 2$ \cite{NKG,Forest} --
multiplied by 1/3, which accounts for the
relative abundance of $pp$ pairs in the two-nucleon short-range correlation.
The integration of the neutron momentum up to 100 MeV/c leads to an
additional factor of 1/2.
Thus, these estimations yield an overall factor of $\approx 1/3$
by which $\sigma (\gamma (pp) \to pp)$ should be scaled 
in order for it to correspond to the $\gamma \; ^3{\rm He} \to pp + n$
cross section.
The overall factor of $1/3$ is a conservative estimate; the inclusion 
of three-nucleon correlations in $^3$He would increase this factor.
Thus, in the RNA approach, 
$d\sigma(\gamma \; ^3{\rm He} \to pp + n)$ $/$ 
$d\sigma(\gamma d \to p n)$ $=$ $16/3$.
 
\medskip

\noindent{\bf TQC model:} 
Estimates for the $\sigma (\gamma (pp) \to pp)$ to
$\sigma(\gamma (pn) \to pn)$ cross-section ratio 
in the TQC model are 
underway \cite{radyushkinmodel}.
We expect the same $^3$He correction
factor of 1/3 that we apply to the RNA model.

\medskip

\noindent{\bf QGS model:} 
In the QGS model, since the break-up cross section is defined by 
the effective Regge trajectory, we would expect the Regge
trajectories to be similar, so the
$\sigma(\gamma (pn) \to pn)$ and $\sigma (\gamma (pp) \to pp)$ 
cross sections are of similar magnitude \cite{pcom}.
We assume that this is multiplied by the same $^3$He correction
factor of 1/3 that we apply to the RNA model.

\medskip

\noindent{\bf HRM model:} 
The differential cross section within the HRM model is \cite{tobepre}: 
$$
\!\!\!\!\!
{d\sigma\over dt d^3p_n} = 
 \left({14\over 15}\right)^2 
{8\pi^4\alpha_{EM} \over s-M^2_{^3{\rm He}}} 
\;\; {d\sigma^{pp}(s_{pp},t_{N})\over dt}
\times
$$
\begin{equation}
\;\;\;\;\;{1\over 2}
\left|\sum\limits_{spins}\int \Psi^{^3{\rm He}}(p_1,p_2,p_n)\sqrt{M_N}{d^2p_{2T}\over 
(2\pi)^2}\right|^2,
\label{eq:3hecsformula}
\end{equation}
where $s=(P_\gamma + P_{^3{\rm He}})^2$, $t=(P_p-P_\gamma)$
$s_{pp}=(P_\gamma + P_{^3{\rm He}}-P_n)^2$, and 
$t_{N}\approx {1\over 2}t$. 
The $pp$ elastic cross section is ${d\sigma^{pp}/dt}$.
The momentum of the recoil neutron is $p_n$.
In the argument of the $^3$He nuclear wave function,
$\vec p_1 = -\vec p_2 - \vec p_n$ and 
$p_{1z} \approx p_{2z} \approx -{p_{nz}\over 2}$ near 90$^{\circ}$. 
The $pp$ scattering cross section was obtained from a fit to the existing $pp$
data \cite{ppd}. 
The overall factor $({14\over 15})$ is obtained based on the 
quark-interchange model of hard NN scattering utilizing the $SU(6)$ 
wave function of nucleons. 
This introduces an uncertainty in the estimates of the cross section
at the level of 10 -- 20\%.
The  $^3$He wave function is that of   
Ref.\cite{NKG}, obtained  by solving the Faddeev equation with a realistic 
NN potential. 
The predicted cross section is made singly differential
by integrating over neutron momentum, up to 100 MeV/c.

\medskip

\begin{figure}[th]
\centerline{\epsfig{angle=0,width=7.75cm,file=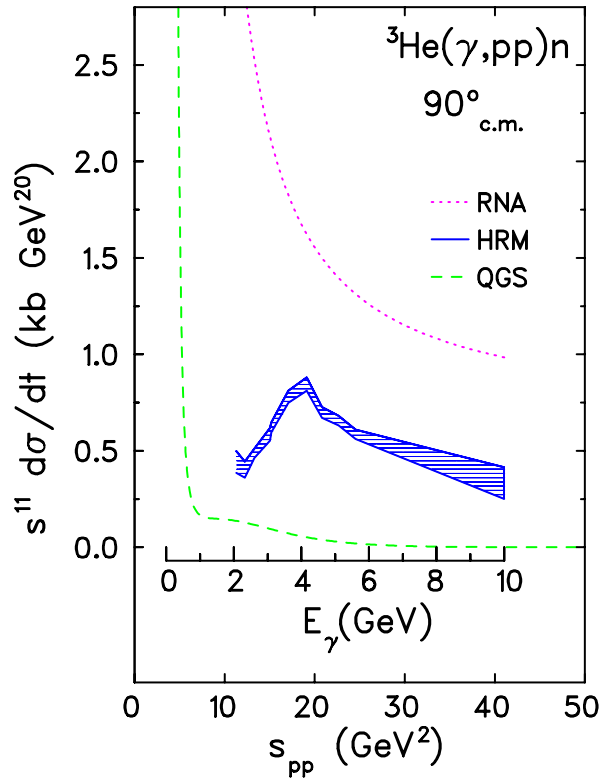} }
\vspace*{0.5mm}
\caption{\it Predictions for $\gamma \; ^3$He$ \to p p + n$ at 
$\theta_{\rm c.m.}$ $=$ 90$\, ^{\circ}$. 
The line types are the same as for Figure 1.
The horizontal scale is $s$ for the $\gamma p p$ system;
the photon energy scale is also shown.
\label{gheppn}}
\end{figure}

Figure~\ref{gheppn} shows predictions based on the models considered above 
for $90^{\circ}$ two-body break-up kinematics. 
The $\gamma \; ^3$He$  \to p p + n$ 
cross section has been integrated over the neutron momentum up to 100 MeV/c.

These predictions ignore nuclear corrections due to the soft 
rescattering of the nucleons in the final state. 
We argue here that they are only small corrections in the 
kinematics discussed.
For energetic protons rescattering on the slow spectator neutron, 
the mean squared value of the 
momentum transferred during the soft rescattering is
200 -- 250 MeV$^2$/c$^2$. 
Restricting the neutron momenta to $\le$ 100 MeV/c significantly
reduces the soft-rescattering. 
This effect can be reliably calculated 
within the eikonal approximation. Preliminary estimates yield 
5 -- 10 \% corrections in the range of 40 -- 90$^{\circ}$ c.m.\ angles.

Another correction is due to primary reactions on
the $pn$ pair, with subsequent soft $pn \to np$ charge-exchange 
rescattering of the energetic neutron with the slow spectator proton.
In the energy range of this study, 
the charge-exchange soft rescattering is suppressed by a factor of $1/s$ 
as compared to the non-charge-exchange soft rescattering, and results in 
only a 1 -- 2 \% correction. This estimate takes into account
the larger probability of $pn$ than $pp$ pairs in $^3{\rm He}$.

It is important to note that the models considered above predict a sizable 
cross section for the break up of the $pp$ pair,
larger than that for the $pn$ pair, for two of the three models shown.
This prediction is rather striking since 
at low energies it is well known \cite{legs} that 
photodisintegration of the $pp$ system is suppressed  as compared to 
$pn$.

Within a mesonic description of the interaction, the 90$^{\circ}$ 
break up of a $pp$ 
pair will  be significantly suppressed as compared to $pn$ since for the  
$pp$ pair the exchanged mesons are neutral and do not couple to the photon. 
In a quark-gluon picture, the exchanged particles are quarks, and
the suppression will be absent. 
As a result an  experimental observation of a larger cross section for the 
$pp$ break-up reaction will be an indication of the dominance of quark-gluon 
dynamics in the reaction.

\section{Oscillations with Energy}

The possibility that the final-state high-$p_T$ proton pair is formed due to 
the hard interaction of the two outgoing protons might 
produce energy oscillations, as seen in the $pp$ cross section. 
The quark counting rule predicts $\frac{d\sigma}{dt} \sim s^{-10}$
for high-energy, large-angle $pp \to pp$ elastic scattering.
The $pp$ elastic data are globally consistent over a large number of decades
with the power law  \cite{ppd,kn:it4}. 
However, it was already noted in 1974 \cite{kn:hend} 
that a more detailed examination of the data indicated significant 
deviations from scaling. The deviations  are known as ``oscillations''
and were interpreted as resulting from interference between the pQCD amplitude
and an additional nonperturbative component.

Ralston and Pire \cite{kn:rals82} suggested that 
the interference is between a small size configuration pQCD scattering 
and an independent scattering of {\bf all} 
valence quarks discussed by Landshoff \cite{kn:lfmult},
governed by the so-called chromo-Coulomb phase.
Brodsky and deTeramond \cite{kn:bdte} suggested that the oscillations
are due to the presence of two broad resonances (or threshold enhancements) 
which interfere with the  standard pQCD amplitude.
For a review of wide-angle processes, see \cite{kn:intr}.

\begin{figure}[th]
\centerline{\epsfig{angle=0,width=7.75cm,file=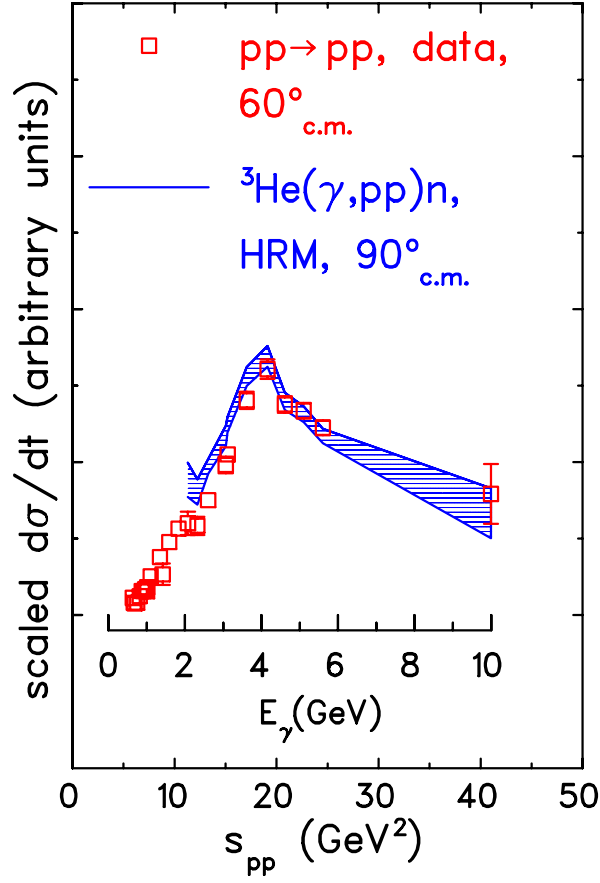}}
\vspace*{0.1mm} 
\caption{\it Energy dependence of the  
$\gamma \; ^3$He$ \to p p + n$ cross section predictions multiplied 
by $s^{11}$, compared with the energy dependence of 
the $pp \to pp$ cross section multiplied by $s^{10}$ and rescaled by an overall
constant, to emphasize the similarity in the energy dependences.
The horizontal scale is $s$ for the $\gamma p p$ and $pp$ systems;
the photon energy scale is also shown.
The different angles for the two reactions are chosen to match the momentum
transfers, as discussed in the text.
The shaded band is the HRM result, which is based on the $pp$ elastic data. 
}
\label{gheppn_os}
\end{figure}

Whatever is the correct interpretation of the oscillation, if
the hard two-body break-up reaction proceeds through the 
hard interaction of two protons, similar oscillations 
could be seen in the $\gamma \; ^3{\rm He} \to pp +n$ cross 
section, normalized by a factor of $s^{11}$, as a function of the incident 
photon energy, in the same region of $s$ where $pp$ oscillations are observed.
Figure~\ref{gheppn_os} compares the 
energy dependence of $pp$ cross section  with that of 
$\gamma \; ^3{\rm He} \to pp +n$ cross section at $90^0$ $\gamma - (pp)$ 
center of mass scattering ($-{t\over s_{pp}}\approx {1\over 2}$), 
calculated within the HRM model, which assumes the dominance of the  
contribution of hard $pp$ rescattering in the photodisintegration reaction.
Note  that according to Eq.(\ref{eq:3hecsformula}) the $pp$ cross section  
that enters in the $\gamma +^3{\rm He}\to pp+n$ cross section is defined at 
$s_{pp}$ and $t_{N}\approx {t\over 2}$. 
As a result, in Figure~\ref{gheppn_os} one 
compares with $pp$ cross sections defined at $\approx 60^0$ 
($-{t_N\over s}\approx {1\over 4}$) \cite{ppd}.
In contrast to the situation displayed in Figure~\ref{gheppn_os},
the precision of the $pn$ and the $\gamma d \to p n$ data is
insufficient to show if oscillations are indeed present for those reactions.

Brodsky and de Teramond \cite{kn:bdte} suggested that the oscillations and 
also the associated large values of the $A_{NN}$ spin correlations observed in 
$pp$ scattering \cite{Court:1986dh} are due to the presence of broad 
resonances associated with the onset of the strangeness and charm 
thresholds in the intermediate state of the $pp \to pp$ amplitude.
If this is correct, then one would also expect to see similarly strong 
spin-spin correlations in the emerging proton pair at the corresponding 
invariant mass. The observation of the large cross sections predicted
here then leads to important related polarization measurements.
One also would expect the production of doubly-charged 
final states with baryon number $B=2$ containing charmed hadrons
in $\gamma He^3 \to n X$ at missing mass $m_X > 5 $ GeV.
The threshold for open charm production is 
$\sqrt{s} > 6 $ GeV, 
$E^\gamma_{lab}  = {s-M^2_{He^3}\over 2 M_{He^3}} > 4.5 $ GeV.

\section{The $\alpha_n$ Distribution}
The recoil neutron in $\gamma \; ^3{\rm He} \to p p + n$ gives 
an additional degree of freedom for checking the underlying mechanism of  
hard $pp$ pair production. The observable which is best suited for this 
purpose is the light-cone momentum distribution of the recoil
neutron, defined as a function of $\alpha_n = {E_n - p_n^z\over m_{^3He}/3}$.
We use here light-cone variables in which the $\alpha$'s are defined 
as follows:
\begin{equation}
\alpha = 
A{{E^N-p^N_z}\over{E^A-p^A_z}} \approx  {E_N-p^N_z \over m_N},
\end{equation}
where the $z$ direction is chosen in the direction of the incident 
photon beam.

With the above definitions, $\alpha$ for the incident photon is 
exactly zero, while $\alpha$ for the $^3$He target is 3.
Conservation of $\alpha$ allows $\alpha_n$ to be determined
from the measurement of the light-cone fractions of the protons:
\begin{equation}
\alpha_{\gamma}+\alpha_{^3{\rm He}} = 0+3 = 
\alpha_{p_1}+\alpha_{p_2}+\alpha_{n}.
\label{eq:exeq1}
\end{equation}
Therefore:
\begin{equation}
\alpha_{n} = 3- \alpha_{p_1} - \alpha_{p_2}.
\label{eq:exeq2}
\end{equation}
An important feature of high-energy small-angle final-state
rescattering is that it does not change the light-cone fractions of
the fast protons -- see e.g. \cite{MS01}. As a result,
the experimentally determined 
$\alpha_n$ coincides with the value of $\alpha_n$ in the
initial state and unambiguously measures the 
light-cone fraction of the two-proton subsystem in the
$^3$He wave function. Furthermore, in the $^3$He wave function 
the c.m.\ momentum distribution of the $NN$ pair depends on 
the relative momentum of the nucleons in the pair, so one can probe the magnitude 
of the momentum in the $pp$ pair involved in the hard disintegration.

\begin{figure}[th]
\centerline{\epsfig{angle=0,width=7.75cm,file=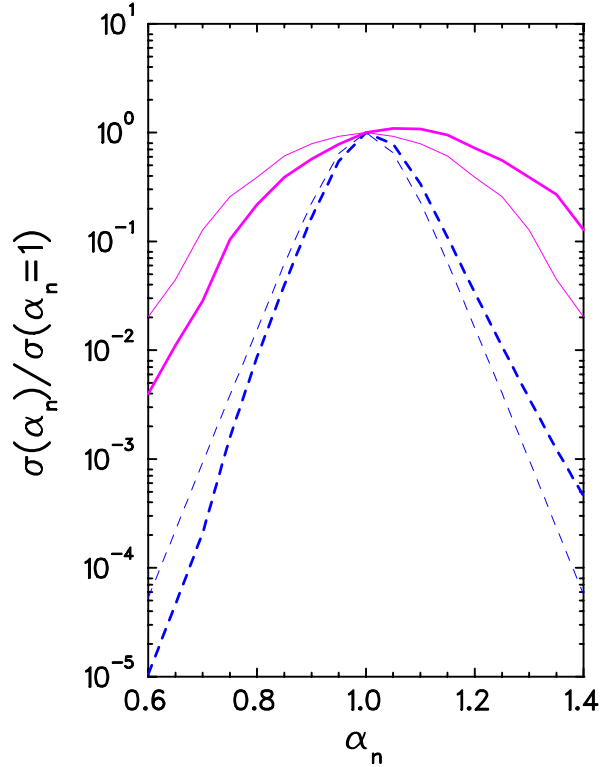}}
\vspace*{0.1mm} 
\caption{\it The $\alpha_n$ dependence of the  
$\gamma \; ^3$He$ \to p p + n$ cross section 
calculated within RNA (bold solid line) and HRM (bold dashed line) 
models. $\sigma(\alpha_n)$ corresponds to the differential cross section 
scaled by $s_{pp}^{11}$. Thin solid and dashed lines correspond to the same 
calculations scaled by $s_d^{11}$. All calculations are normalized 
to one at $\alpha_n=1$.}
\label{fig:alphan}
\end{figure}

To illustrate the sensitivity of the $\alpha_n$ distribution to the mechanism 
of the high-$p_T$ disintegration of a $pp$ pair, we compare in
Fig.~\ref{fig:alphan} 
the $\alpha_n$  dependence of the differential cross section 
${d\sigma\over dt d^2p_T d\alpha_n/\alpha_n}$ calculated in the framework of
the RNA and HRM models. The calculations are done for a
scattering of two protons in the final state at fixed initial
photon energy $E_{\gamma}$ $=$ 4~GeV and $\theta_{\rm c.m.}$ $=$ 90$^{\circ}$. 
Within the RNA approximation (solid line),
the $\alpha_n$ distribution is calculated for configurations in 
which the relative transverse momentum of the $pp$ pair 
is equal to the transverse momentum of the final 
protons $p_T\sim GeV/c$.   The estimate within the HRM model
is done using Eq.(\ref{eq:3hecsformula}). In the latter case, the 
internal momenta in the $pp$ pair contributing to the cross section 
are $\le$ 300 MeV/c.
The results presented in Fig.~\ref{fig:alphan} provide
substantially different predictions for the $\alpha_n$ distribution.
Qualitatively, the much broader distribution of $\alpha_n$ in the RNA
model is due to selection of large momenta of protons in the
$^3He$ wave function, which leads
to a broader distribution of neutron momenta.

Another feature of the $\alpha_n$ distribution is 
that the strong $s$ dependence, $\sim$ $s^{-11}$, of the hard disintegration 
cross section  will tend to suppress (increase) the contribution from those
values of $\alpha_n$  which increase (decrease) the effective
$s_{pp}\approx 2E_{\gamma}M_d{3-\alpha_n\over 2}+M_d^2$ involved 
in the $\gamma + pp$ subprocess.  As a result one expects the  
$\alpha$ distribution to be asymmetric about $\alpha_n=1$. The extent of 
the asymmetry depends strongly on the exponent in the $s$ dependence of
hard disintegration cross section. 
To illustrate this phenomenon, in Fig.~\ref{fig:alphan} we 
compare the $\alpha_n$ distributions within the RNA and HRM models, rescaled
in one case by $s_{pp}^{11}$ (bold solid and dashed lines)
and in other case  by $s_{d}^{11}$ ($s_d = 2E_{\gamma}M_d + M_d^2$)
(thin lines). This comparison demonstrates that the measurement of the 
$\alpha_n$ asymmetry will give us an additional tool in verifying the 
energy ($s$) dependence of the disintegration cross section.

\section{Experiments}
Data for $^3$He photodisintegration have already been obtained
by the CLAS collaboration, up to energies of 1.5 GeV, but no results
are available as yet \cite{bermanpc}.
As the onset of scaling in deuteron photodisintegration is just above 1 GeV,
it will be interesting to see if there is a similar onset for $^3$He,
and, if so, what is the ratio of $^3$He to deuteron photodisintegration
cross sections.

Studying the $\gamma \; ^3{\rm He} \to pp + n$ reaction to significantly 
higher energies requires measuring a small cross section reaction that
generates two high transverse momentum protons.
It is only possible in Hall A of the Thomas Jefferson National 
Accelerator Facility using Bremsstrahlung photons, produced by 
the electron beam passing through a photon radiator.
The maximum energy of the Bremsstrahlung beam is essentially 
equal to the incident electron kinetic energy. 
The two outgoing protons,
each with about half the incident beam energy, can be detected in 
coincidence with the two existing high resolution spectrometers (HRS).
The energy dependence of the differential cross section
for $\theta_{\rm c.m.}$ $\approx$ 90$^{\circ}$
can be measured up to $E_{\gamma}$ $\approx$ 5 GeV
with the existing equipment, if the
cross sections are as large as predicted here.
In contrast, it has only been possible to measure
deuteron photodisintegration up to 4 GeV at 
$\theta_{\rm c.m.}$ $\approx$ 90$^{\circ}$, due to the
rapid decrease of its cross section.
If the large predicted cross sections are verified, polarization
measurements will be possible to $\approx$ 4 GeV.
A  measurement of $A_{NN}$ of the two outgoing protons would be 
particularly interesting in view of the observed dramatic spin effects in 
elastic $pp \to pp$, and will require a dedicated measurement with 
polarimeters in both spectrometers.
With the proposed 12 GeV upgrade, including the proposed higher momentum
spectrometer for Hall A, it would be possible to extend the measurements
up to about 7 GeV in a matter of weeks, limited by the
maximum momentum in the HRS spectrometer.

\section{Summary and Outlook}
A unique signature of quark-gluon degrees of freedom in 
hard photodisintegration reactions is the prediction of a sizable  
cross section, larger for $pp$ than for $pn$ pairs.
If the hard photodisintegration process can be factorized
so that it depends on the $NN$ scattering amplitude, then
the oscillations apparent in $pp$ scattering could 
be reflected in the measured cross sections.
Comparing the predictions presented here to data 
could put our understanding of deuteron
photodisintegration on a firmer basis, and would be a significant step
toward a general understanding of hard nuclear photo-reactions
at intermediate energies.

The observation of oscillations with energy 
would give us a new tool in studies of color coherence phenomena in hard
nuclear reactions. 
The investigation of $A$ dependence of the reaction 
extended to nuclei with $A>3$ would 
allow a study of the nature of these oscillations. 
For instance, if the oscillations are the result of the interplay of 
soft and hard scattering amplitudes, one expects more absorption for 
the soft part of the total amplitude -- 
a phenomenon known as a nuclear filtering.

We also observe that determining the shape and the asymmetry of the $\alpha_n$ 
distribution in the hard  $\gamma \; ^3He \rightarrow pp + n$  reaction gives 
an additional experimental tool in studying the  dynamics  of the high energy 
disintegration of a $NN$ pair.

\medskip
\medskip

We thank A.~Radyushkin and L.~Kondratyuk for useful discussions.\\
The Stanford Linear Accelerator (supporting in part SJB) is funded by the
Department of Energy under contract number DE-AC03-76SF00515.
The Southeastern Universities Research Association operates
the Thomas Jefferson National Accelerator Facility 
(supporting in part SJB, RG and MS)
under U.S.\ DOE contract DE-AC05-84ER40150. 
RG acknowledges the support of the U.S.\ National Science Foundation, 
grant PHY-00-98642.
JRH acknowledges the support of the U.S.\ DOE under contract
DE-FG02-98ER41087. GAM acknowledges the support of the U.S.\ DOE under contract
DE-FG03-97ER41014.
EP is supported by the Israel Science Foundation founded by the Israel 
Academy of Science and Humanities. 
MS acknowledges the support of the U.S.\ DOE under contract DE-FG02-01ER-41172.

\end{document}